\documentstyle[12pt,epsfig]{article}
\begin{document}
\begin{center}
\vspace{1.5in}
{\Large Liquid-Drop and Independent-Particle Models Revisited}
\end{center}
\vspace{.4in}
\begin{center}
{\bf S. Afsar Abbas}\\
Centre for Theoretical Physics\\ 
JMI, New Delhi - 110025, India\\
\vspace{.1in}
email: afsar.ctp@jmi.ac.in
\end{center}
\vspace{.5in}
\begin{center}
{\bf Abstract }
\end{center}
\vspace{.3in}

Traditionally, the difference in binding energy from the experimental 
value with respect to the theoretical liquid-drop model value, 
has been seen as indication of independent-particle character 
along with magicity for particular number of protons and neutrons. 
We study this carefully to 
demonstrate that it actually indicates that the liquid-drop and the 
independent-particle phases of the nucleus have equal fundamental 
primacy and coexist simultaneously in a nucleus to provide
a complete and a consistent description of the same.

\newpage

Since the last seven decades or so,  
the nuclear theory has had a dual approach
- one "microscopic" which centres around the single particle 
characteristics of a nucleon in a nucleus 
(call it Independent-Particle Model - IPM) 
and the other is "macroscopic" and which is concerned with the gross 
properties of the nucleus (call it Liquid-Drop Model - LDM).
It is commonly believed that as the IPM is quantum mechanical in 
character, it is more basic than the LDM which is classical in character.
So within this canonical perspective, 
IPM is more general and LDM is less so. 

There are several properties of nuclei which are indicative of its
IPM character. One of the most quoted one is a graph showing the 
difference between the theoretical liquid-drop value and the experimental 
value of the binding energy of nuclei. This clearly shows the shell 
characteristics as well as magicities in nuclei. The graph
from Ref. [1] and Ref. [2, p.31] is shown here in Fig. 1.

To quote Norman Cook [2, p.31], "The most unambiguous indication of nuclear 
shell structure comes from the data on total binding energies. By 
calculating the expected binding energy using the liquid-drop model 
(without shell corrections) and subtracting it from the experimental 
value, the deviation from simple liquid-drop conception can be determined.
Relatively large deviations are obtained at Z=28, 50 and 82
and N=28,50, 82 and 126. The deviations are indications of slightly higher
binding energies for nuclei with these number of protons and neutrons,
and this supports the idea of relatively tightly bound, 
compact closed shells." 

The semi-empirical mass formula as given in Ref. [1] is



\vspace{.2in}

$M(Th)_{Total} = M(Th)_{LDM} + M(Th)_{shell-correction}$

\vspace{.2in}


They first did the four parameter liquid-drop model fitting very 
carefully.
We already know that even the primitive formula of Weizsacker-Bethe with 
just the liquid-drop part does very well, indeed. In the case of the much 
improved fitting by Myer and Swiatecki, 
only within the liquid-drop framework, 
the fits become much better. They also point out that the liquid-drop 
part was fitted remarkably well [1].
Also as stated [1. p.28], " .... because the liquid-drop parameters govern 
the overall trends and the shell-correction parameters govern local
irregularities, with the result that one may fit four of the parameters
( {\it of the liquid-drop part} )
almost independently of the remaining three
( {\it of the shell-correction part} ) ...."
( Note: italics insertions are mine ).
Hence actually -


\vspace{.2in}

$M(Th)_{LDM} \approx M(expt)_{LDM}$

\vspace{.2in}


Hence one finds that 


\vspace{.2in}

$M(expt)_{Total} - M(expt)_{LDM}  = M(expt)_{shell-correction}$

\vspace{.2in}


So what is plotted in Fig. 1 
is the "experimentally" determined shell-correction terms which are 
IPM-like in character.

What is the above expression trying to tell us?
To understand this we plot Fig. 2 from Ref. [2, p.24].
This figure displays boiling points, the temperature at which the liquid 
to gas transition occurs for various elements (in degrees Centigrade).
Clearly the inert gases show the lowest boiling points for a row in the 
periodic table and with higher boiling points for elements in between.
So the liquid-inert-atoms find it easiest to convert to the gaseous phase.

Now let us compare the two figures. We are struck by a remarkable 
similarity between the two. 
{\bf Fig. 1 is nothing but the boiling point curve for various nuclei}. 
Clearly there are two phases in a nucleus -
the liquid phase and the gaseous phase. 
Exactly as for the atoms, nuclei with
magic numbers find it easiest to go from the liquid phase to the gaseous
phase. The nuclei in between, gradually find it harder to convert from the 
liquid to the gaseous phase and hence this occurs at a higher temperature 
- exactly as in the corresponding atomic case.

The energy scale can be treated as the temperature 
scale in the boiling point curve (Fig. 1). 
This is the experimentally determined shell-correction terms, which are 
IPM-like in character. 
Now we can treat the IPM phase as the gaseous phase [2]. 
So both the shell-corrections  ( which is providing 
a temperature scale here ) and the IPM of the nucleus are gaseous 
in nature.
 
This is good, as in the classical thermodynamics, the "temperature" is
measure of the average kinetic energy of a molecule in a 
gas. This is microscopic definition of a quantity which also manifests 
itself in entirely macroscopic frameworks as well.
So here in our case too, we find that microscopic shell-corrections act as
temperature scale determining the boiling point curve of the nuclei.
This parallelism between the way temperature is defined in classical 
thermodynamics and the way that it is defined in a nucleus, attests to
the consistency of our interpretation.

However note that this is true 
for different nuclei which are bound. So the scale is "internal" 
temperature at which both the "liquid" drop characteristic phase and the 
"gaseous" characteristic phase exit in 
the same nucleus simultaneously.

This is an internal temperature at which the two phases co-exist. 
Just as at 100 degrees Centigrade the liquid and the gas phases of water
coexist at STP, so too in a nucleus, due to an internal temperature 
created and which is specific for a particular nucleus, 
the liquid phase (LDM) and the gas phases (IPM) co-exit simultaneously.

This is a clear and unambiguous demonstration of the fact that both the 
liquid-drop character and the gas-like (IPM) character are equally real. 
Each is as fundamental as the other. 
Also for a particular nucleus, these two phases co-exist simultaneously. 
Neither of these can be transformed away as this is dependent upon 
an internal temperature scale which itself arise from an irreducible 
interplay of the two phases - the LDM an the IPM.

This co-existence of the LDM and the IPM phases explains as to why 
both of these two conflicting approaches have been pretty successful in 
describing nuclei.

This paper gives further perspicuous support to the recent paper by the 
author [Ref. 3], wherein a new fundamental duality in nuclei was 
discussed. This duality is the co-existence  of classical 
liquid-drop characteristics along with the quantum mechanical 
independent-particle characteristics in a nucleus.

\newpage

\vspace{.9in}

{\bf References} 

\vspace{.9in}

1. W. D. Myers and W. J. Swiatecki, {\it Nucl. Phys.} {\bf 81} (1966) 1

\vspace{.5in}

2. Norman D. Cook, {\it "Models of the Atomic Nucleus"}, 
   Springer Verlag, Berlin, {\bf 2006}

\vspace{.5in}

3. S. Afsar Abbas, {\it "A new fundamental duality in nuclei and its 
   implications for quantum mechanics"}, {\bf arXiv:0811.0435}

\newpage

\begin{figure}
\caption{ The difference between the experimental masses and the
theoretical liquid-drop masses (Refs. [1 and 2 p.31]) }
\epsfclipon
\epsfxsize=0.99\textwidth
\epsfbox{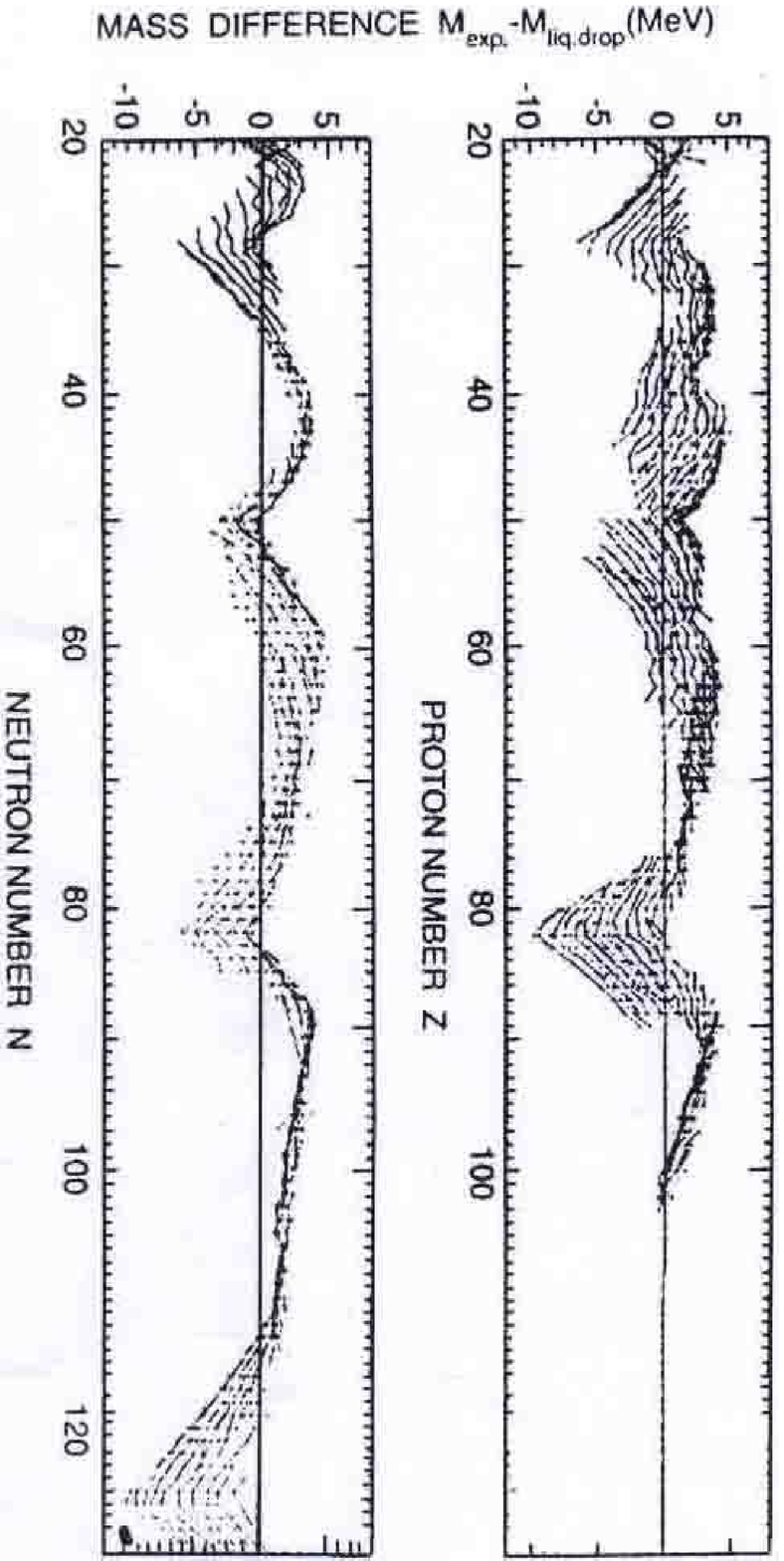}
\end{figure}

\vspace{.4in}

\begin{figure}
\caption{ Boiling points of the various elements (in degrees 
Centigrade) (Ref. [2, p.24)] } 
\epsfclipon
\epsfxsize=0.99\textwidth
\epsfbox{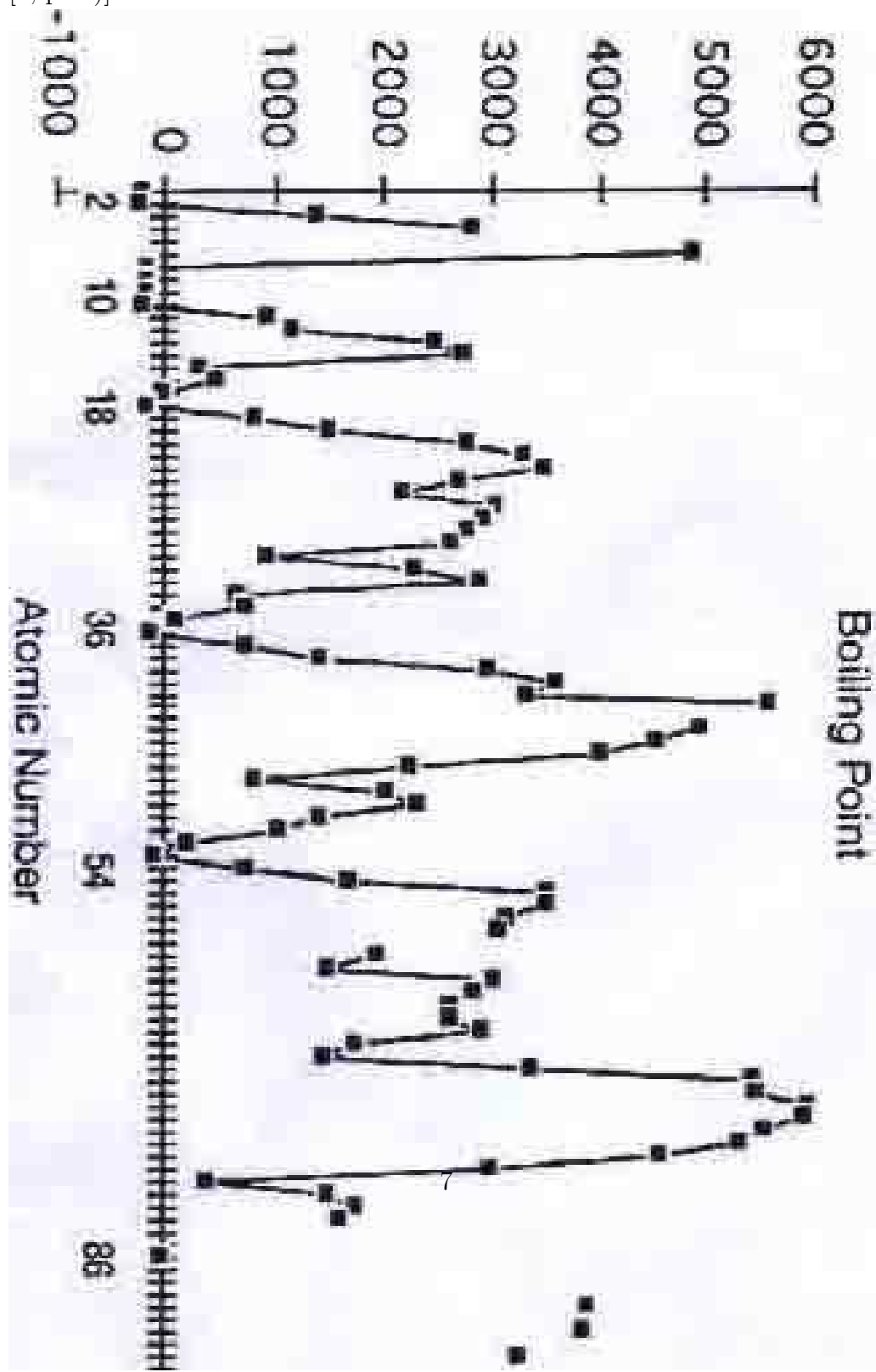}
\end{figure}

\end{document}